\documentclass[showpacs,aps,pre,preprint]{revtex4}
\usepackage{graphicx}
\usepackage{amsfonts}

\newcommand{\be}{\begin{equation}}
\newcommand{\ee}{\end{equation}}
\newcommand{\bea}{\begin{eqnarray}}
\newcommand{\eea}{\end{eqnarray}}
\newcommand{\sn}{{\rm sn}}

\newcommand{\dn}{{\rm dn}}
\newcommand{\cn}{{\rm cn}}

\def\eps{\varepsilon}

\begin{document}

\title{Solitons in strongly driven discrete nonlinear Schr\"odinger-type   models}

\author{Josselin Garnier$^1$ \footnote[2]{Corresponding author
(garnier@math.jussieu.fr)}, Fatkhulla Kh. Abdullaev$^2$ ,
and Mario Salerno$^3$} \affiliation{$^1$ Laboratoire de
Probabilit\'es et Mod\`eles Al\'eatoires \& Laboratoire
Jacques-Louis Lions, Universit{\'e}
Paris VII, 2 Place Jussieu, 75251 Paris Cedex 5, France\\
$^2$ Physical-Technical Institute of the Uzbekistan Academy of
Sciences, 700084, Tashkent-84, G.Mavlyanov str.,2-b, Uzbekistan\\
$^3$ Dipartimento di Fisica "E.R. Caianiello", Universit\'a di
Salerno, 84081 Baronissi (SA), Italy}

\begin{abstract}
Discrete solitons in the Ablowitz-Ladik (AL) and discrete
nonlinear Schr\"odinger (DNLS) equations with damping and strong
rapid drive are investigated. The averaged equations have the
forms of the parametric AL and DNLS equations. A new type of
parametric bright discrete soliton and cnoidal waves are found and
the stability properties are analyzed. The analytical predictions
of the perturbed inverse scattering transform are confirmed by the
numerical simulations of the AL and DNLS equations with rapidly
varying drive and damping.
\end{abstract}
\maketitle

\section{Introduction}

Recently the problem of dynamics of nonlinear lattices under
strong and rapid modulations of parameters has attracted a lot of attention.
Two systems have been analyzed. The first one is the diffraction-managed
array of optical waveguides, with diffraction varying periodically
 along the beam propagation \cite{Abl1}. The model
is described by the discrete nonlinear Schr\"odinger (DNSL) equation with
rapidly and strongly varying in time tunnel-coupling between sites coefficient $c(t)$:
\begin{equation}\label{dnls1}
i {u_{n}}_{t} + \frac{1}{\epsilon}c(\frac{t}{\epsilon})(u_{n+1}+u_{n-1})
+ 2 |u_{n}|^2 u_n =0,
\end{equation}
where $\epsilon \ll 1$. The analysis exhibits the existence of a
new type of discrete spatial optical solitons with beam width and
peak amplitude evolving periodically during propagation.

The second system is the Bose-Einstein condensate in a periodic
(in space) potential with a varying (in time) scattering length.
In the tight-binding approximation this system is described by the
DNLS equation \cite{smerzi-abdul-01} with strongly and rapidly
varying in time nonlinearity coefficient $\kappa(t)$:
\begin{equation}\label{dnls2}
i {u_{n}}_{t} + (u_{n+1}+u_{n-1})
+ \frac{1}{\epsilon}\kappa(\frac{t}{\epsilon})|u_{n}|^2 u_n =0.
\end{equation}
It was shown that this system supports  nonlinearity-managed
discrete solitons \cite{Abd1}. In a more general context it is of
interest to investigate the influence of rapid perturbations  on
the dynamics of discrete solitons in nonlinear lattices.
The case of strongly and rapidly varying external drivers
is particularly important for applications. This problem is encountered both
in the study of the dynamics of a magnetic flux quantum in an
array of long Josephson junctions with varying ac current
\cite{Ust} and in the evolution of an optical field in a nonlinear
chain of resonators or microcavities in presence of pumping
\cite{Christ,Stal,Pesh,Gorb}.

In this paper we consider the influence of a rapid strong drive on
discrete bright solitons and cnoidal waves of  the Ablowitz-Ladik
(AL) and the DNLS equations with damping. Although the AL system
has scarce physical applications it has many advantages from the
analytical point view such as the complete integrability of the
unperturbed system, existence of moving discrete solitons etc. In
some regions of the parameter space, the DNLS equation can be
described as a perturbation of the AL model, a feature that we
shall take advantage of in the following. The dynamics of discrete
solitons in AL and DNLS equations under the influence of damping
and slowly varying driving field has been studied in
Refs.~\cite{Bount,Hennig}. Recently the influence of parametric
drivers on the stability of strongly localized modes of the  DNLS
equation near the anti-continuum limit has been investigated in
\cite{Kevr1}.
The stability of solitons in the
continuous parametrically driven NLS equation
has been studied in \cite{Bar1,Bar2}.
Here we address a general discrete nonlinear
system with a strong rapid drive modeled by the following
equation:
\begin{equation}
\label{eq:al0} i {u_n}_t +  [ u_{n+1}+u_{n-1} - (2 +\omega)u_n] +
|u_n|^2 [ (1-\chi) (u_{n+1} +u_{n-1}) +2 \chi u_n] =
\frac{1}{\eps} f(\frac{t}{\eps}) - i \gamma u_n,
\end{equation}
where $f$ is a zero-mean periodic function with period $1$ that
describes the rapid drive, and the small parameter $\eps$ is the
period of the drive. Here the parameter $\gamma \geq 0$ denotes
the damping term, $\omega$ is the propagation constant in
optics (chemical potential in BECs), and the parameter $\chi \in
[0,1]$ characterizes the type of nonlinearity. For $\chi=0$ the
nonlinearity is of intersite type as in the AL model while for
$\chi=1$ we have the onsite nonlinearity of the DNLS equation. In
absence of strong rapid drive and damping Eq.~(\ref{eq:al0})
coincides with the Salerno model \cite{Salerno92} which
interpolates between the AL model and the DNLS model. We show that
by averaging out the fast time scale one can reduce Eq.
(\ref{eq:al0}), for particular choices of parameters, to the
parametrically driven AL and DNLS equations with damping. The
existence of parametric discrete bright solitons and cnoidal waves
of these equations is then investigated and the stability
properties is analyzed both analytically and numerically. We find
that the analytical predictions obtained from the averaged
equations by means of a perturbation scheme  based on the inverse
scattering method are in good agreement with direct numerical
simulations of the problem with rapidly varying drive and damping.

Finally, we remark that the physical systems where parametric
discrete solitons of the type considered in this paper can be
realized are chains of linearly coupled nonlinear
microcavities \cite{Efr2} and nonlinear waveguide arrays with
dielectric mirror at the ends, driven by an external
time-dependent field \cite{Pesh,Gorb}. In this case the equation
describing the discrete cavity solitons has the form of the parametric
DNLSE
\begin{equation}
i {u_{n}}_{t} + \omega u_{n} + i\gamma u_{n} + \alpha |u_{n}|^2 u_{n} +
C(u_{n+1} + u_{n-1} - 2u_{n}) = F_{n}(t),
\end{equation}
where $\omega$ is the detuning from the linear resonance
parameter, $\gamma$ is the effective damping parameter, $C$ is the
effective coupling between adjacent waveguides, $F_{n}(t)$ is the
input field in the $n$-th waveguide. Injecting a homogeneous
in space and rapidly varying in time field, we can generate
discrete parametric soliton  in this system.

The paper is organized as follows. In Section II we derive
the averaged DNLS-type equation for the strongly and rapidly varying
external drive model. In Section III we analyze the solitons in a
damped AL system with parametric drive. We use the perturbation
theory based on the Inverse Scattering Transform (IST) and study
the stability region of parametric discrete solitons.  Periodic
solutions of the damped AL equation with parametric drive are also
found in this section. The discrete soliton dynamics in
parametrically driven DNLSE is investigated in Section IV.

\section{Averaging}
We look for the solution of Eq.~(\ref{eq:al0}) in the form
\begin{equation}
\label{eq:expand}
u_n(t) = u_n^{(0)}(t,\frac{t}{\eps})+\eps u_n^{(1)}(t,\frac{t}{\eps})+\cdots
\end{equation}
where $u_n^{(0)}$, $u_n^{(1)}$, $\ldots$ are  periodic in the argument $\tau=t/\eps$.
We substitute this ansatz into Eq.~(\ref{eq:al0})
and collect the terms with the same powers
of $\eps$.
We obtain the hierarchy of equations:
\begin{eqnarray*}
&& i {u_n^{(0)}}_\tau =  f(\tau) \\
&& i {u_n^{(0)}}_t + [ u^{(0)}_{n+1}+u^{(0)}_{n-1} - (2 +\omega)u_n^{(0)}]
+
|u_n^{(0)}|^2 [(1-\chi) (u^{(0)}_{n+1} +u^{(0)}_{n-1})
+2\chi u^{(0)}_n] =
 - i \gamma u_n^{(0)} - i {u_n^{(1)}}_\tau
\end{eqnarray*}
The first equation imposes the form of the leading-order term
$$
u_n^{(0)}(t,\tau) = -i F(\tau) + {a}_n(t) \, ,
$$
where $F(\tau)=\int_0^\tau f(s) ds$
is the antiderivative of $f$  and
${a}_n$ depends only on the slow variable $t$.
The second equation is the compatibility equation for the existence
of the expansion (\ref{eq:expand}).
By integrating over a period in $\tau$, we obtain:
\begin{eqnarray*}
i {{a}_n}_t + [ {a}_{n+1}+{a}_{n-1} - (2 +\omega){a}_n]
+ (1-\chi)
 \left< |{a}_n -i F(\cdot)|^2 ({a}_{n+1} +{a}_{n-1}-2i F(\cdot) ) \right>\\
 +2 \chi  \left< |{a}_n -i F(\cdot)|^2 ({a}_n-i F(\cdot) ) \right>
 =
 - i \gamma {a}_n
\end{eqnarray*}
which gives
\begin{eqnarray}
\nonumber
i {{a}_n}_t + [ {a}_{n+1}+{a}_{n-1} - (2
+\omega){a}_n] + |{a}_n |^2 [(1-\chi) ({a}_{n+1} +{a}_{n-1})
+2\chi a_n]\\
 =  - i \gamma {a}_n + \delta \overline{{a}_n} -
\frac{\delta}{2} [(1-\chi) (  {a}_{n+1}+{a}_{n-1}) +2 (1+\chi)
{a}_n],
\label{eq:ala}
\end{eqnarray}
where $\delta=2\left< F^2\right>$.

Thus, the averaging method applied to Eq. (\ref{eq:al0}) leads to
a {\it parametrically driven} nonlinear lattice equation which
reduces to the Salerno model \cite{Salerno92} in absence of
perturbations. It is possible to consider a random forcing instead
than a periodic one. More precisely, we get the same result if the
source $f(\tau)$ is a colored noise with coherence time of order
one. However, the power spectral density of the source should
vanish at zero-frequency. Otherwise it would appear a phase
diffusion and this would destroy the stability of the stationary
solution that we will introduce next.

We introduce the rescaled time ${T}=t (1+\delta(1-\chi)/2)$ and
the rescaled function ${A}_n={a}_n /\sqrt{1+\delta(1-\chi)/2}$.
The averaged equation for the function $A_n(T)$ has the form
\begin{equation}
i {{{A}_n}}_{{T}} + [{A}_{n+1}+{A}_{n-1} - (2 + {\Omega}){A}_n] +
|{A}_n |^2 ({A}_{n+1} +{A}_{n-1}) =  R_n, \label{eq:al1}
\end{equation}
where
\begin{eqnarray*}
\Delta= \frac{\delta}{1+\delta(1-\chi)/2}\, , \ \ \ \
\Gamma= \frac{\gamma}{1+\delta(1-\chi)/2}\, , \ \ \  \
{\Omega}= \frac{\omega-2\delta}{1+\delta(1-\chi)/2}\, ,
\end{eqnarray*}
and the term in the right-hand side is given by
\begin{equation}
\label{eq:pert0} R_n = - i \Gamma A_n + \Delta \overline{A_n}
+\chi  |A_n|^2 ( A_{n+1}+A_{n-1}-2 A_n).
\end{equation}
The left hand side of (\ref{eq:al1}) is the Ablowitz-Ladik (AL)
equation, that is completely integrable. The right-hand side can
be seen as a perturbation of this system. We shall use the
perturbed Inverse Scattering Transform (IST) to study the
evolution dynamics of AL solitons driven by the perturbation
$R_n$.

\section{The damped AL system with parametric drive}
We consider in this section the case $\chi=0$, that is the AL model
with damping and rapid drive.
Therefore, we consider the perturbed AL equation (\ref{eq:al1})
with the perturbation
\begin{equation}
\label{eq:pert1}
R_n = - i \Gamma A_n + \Delta \overline{A_n} .
\end{equation}

\subsection{Perturbed Inverse Scattering Transform}
We assume that the damping parameter
${\gamma}$ and the parametric drive parameter ${\delta}$ are small (but ${\omega}$
can be of order one).
Therefore, $\Gamma$ and $\Delta$ are small, and,
following \cite{vakh1,cai,Dokt}, the evolution
equations for the soliton parameters in the adiabatic approximation
have the form
\begin{eqnarray*}
x_T & =& 2 \frac{\sinh \beta}{\beta} \sin \alpha  + \frac{\sinh\beta}{\beta}
\sum_{n=-\infty}^\infty \frac{(n-x) \cosh\beta(n-x) {\rm Im}(r_n) }{\cosh\beta(n+1-x)
\cosh(\beta(n-1-x) }\;, \\
\beta_T &=& \sinh\beta \sum_{n=-\infty}^\infty \frac{\cosh\beta(n-x) {\rm Im}(r_n) }{\cosh\beta(n+1-x)
\cosh\beta(n-1-x) }\;, \\
\alpha_T &=&   \sinh \beta \sum_{n=-\infty}^\infty \frac{\sinh\beta(n-x) {\rm Re}(r_n) }{\cosh\beta(n+1-x)
\cosh\beta(n-1-x) }\;, \\
\sigma_T &=& 2\cosh \beta \cos \alpha +2 \frac{\sinh \beta}{\beta}\alpha \sin \alpha -2-\Omega +
\sinh \beta
\sum_{n=-\infty}^\infty \frac{(n-x) \sinh\beta(n-x) {\rm Re}(r_n) }{\cosh\beta(n+1-x)
\cosh\beta(n-1-x) } \\
&&-
\cosh \beta
\sum_{n=-\infty}^\infty \frac{\cosh\beta(n-x) {\rm Re}(r_n) }{\cosh\beta(n+1-x)
\cosh\beta(n-1-x) }\\
&& + \alpha \frac{\sinh\beta}{\beta}\sum_{n=-\infty}^\infty
\frac{(n-x) \cosh\beta(n-x) {\rm Im}(r_n) }{\cosh\beta(n+1-x)
\cosh\beta(n-1-x) }.
\end{eqnarray*}
where $r_n=R_n \exp( -i \alpha(n-x)-i\sigma)$.
Using standard analytical tools (Poisson summation formula and
residue theorem), we can compute the right-hand sides of these equations.
The equation for $\beta$ takes the form:
\begin{equation}
\label{eq:al:betaT}
\beta_T = \Delta P^{(\beta)}   +\Gamma G^{(\beta)}
\end{equation}
where
\begin{eqnarray*}
P^{(\beta)} &=& -\sinh^2 \beta \sum_{s=-\infty}^\infty I_\beta(2 \alpha +2\pi s)
\sin(2\pi s x+2\sigma)\;, \\
G^{(\beta)}&=&-2  \tanh \beta
\end{eqnarray*}
with $$
I_\beta(a) = \frac{2\pi \sin a}{\beta \sinh(2\beta) \sinh (\frac{\pi a}{2\beta})}.
$$

The equation for $\alpha$ has the form:
\begin{equation}
\label{eq:al:alphaT}
\alpha_T = \Delta P^{(\alpha)}  +\Gamma G^{(\alpha)}
\end{equation}
where
\begin{eqnarray*}
P^{(\alpha)} &=& -\sinh^2 \beta \sum_{s=-\infty}^\infty K_\beta(2 \alpha +2\pi s)
\sin(2\pi s x+2\sigma)\;, \\
G^{(\alpha)}&=& 0\;,
\end{eqnarray*}
with $$
K_\beta(a) = \frac{2\pi \sin^2( a/2) }{\beta \sinh^2 \beta \sinh (\frac{\pi a}{2\beta})}.
$$

The equation for the soliton center  $x$ has the form:
\begin{equation}
\label{eq:al:xT}
x_T = 2 \frac{\sinh \beta}{\beta} \sin \alpha + \Delta P^{(x)}  +\Gamma G^{(x)}
\end{equation}
where
\begin{eqnarray*}
P^{(x)} &=& - \frac{\sinh^2 \beta}{\beta} \sum_{s=-\infty}^\infty J_\beta(2 \alpha+2\pi s)
\cos(2 \pi s x +2\sigma)\;, \\
G^{(x)} &=& -\frac{\sinh^2 \beta}{\beta} \psi_\beta(x) \;,
\end{eqnarray*}
with $J_\beta(a)= - I_\beta'(a)$ and
$$
\psi_\beta(x) = \sum_{n=-\infty}^\infty \frac{(n-x)
}{\cosh(\beta(n+1-x) \cosh(\beta(n-1-x) } = \frac{4 \pi}{\beta
\sinh(2\beta)} \sum_{s=1}^\infty \frac{\sin(2\pi s x)}{\sinh(
\frac{\pi^2 s}{\beta})}.
$$

Finally, the equation for the soliton phase  $\sigma$ has the form:
\begin{equation}
\label{eq:al:sigmaT}
\sigma_T = 2\cosh \beta \cos \alpha+ 2 \frac{\sinh \beta}{\beta}\alpha \sin \alpha -2
-\Omega
 +
 \Delta  P^{(\sigma)} +\Gamma G^{(\sigma)}
 \end{equation}
where
\begin{eqnarray*}
P^{(\sigma)} &=& -\sinh \beta \sum_{s=-\infty}^\infty
L_\beta(2 \alpha+2\pi s ) \cos( \alpha+ \pi s + 2 \sigma +2\pi s x)\\
&&+\sinh\beta \sum_{s=-\infty}^\infty K_\beta'(2 \alpha+2\pi s)
 \sin(2 \pi s x +2\sigma) \\
&&-\sinh\beta \sum_{s=-\infty}^\infty K_\beta(2 \alpha+2\pi s)
 \cos(2 \pi s x +2\sigma) \\
&&- \frac{\alpha \sinh^2\beta}{2\beta}
\sum_{s=-\infty}^\infty I_\beta'(2 \alpha+2\pi s)
 \sin(2 \pi s x +2\sigma)\;,
\\
G^{(\sigma)} &=& - \alpha \frac{\sinh^2 \beta}{\beta}
\psi_\beta(x)\;,
\end{eqnarray*}
with
$$
L_\beta(a) = \frac{2\pi \sin(a/2)}{\beta \sinh \beta \sinh(\frac{
\pi  a}{2 \beta})}.
$$

\subsection{Parametrically driven AL solitons}
The system (\ref{eq:al:betaT}-\ref{eq:al:sigmaT})
has two fixed points if $\Gamma <\Delta$
(which is equivalent to $\gamma<\delta$)
and $\Omega+\Delta>0$ (which is equivalent to $\omega-\delta>0$):
\begin{eqnarray}
&&\alpha_{\pm}=0\, ,\ \ \ \ \sin (2\sigma_{\pm}) =-\frac{\Gamma}{\Delta}\, , \ \ \ \
\cos(2 \sigma_{\pm} )= \pm \sqrt{1-\frac{\Gamma^2}{\Delta^2}}\, , \\
&& \beta_\pm = {\rm arccosh} \left(1 + \frac{\Omega}{2}
+\frac{\Delta}{2} \cos(2\sigma_\pm) \right).
 \label{eq:defbetapm1}
\end{eqnarray}
In fact, the second point (with subscript $-$)
exists only if $\Omega+\Delta \cos(2\sigma_-)>0$.
The center of the soliton can be arbitrary.
Note that these values correspond to a fixed point of the equation (\ref{eq:al1}).
The soliton $\pm$ is of the form
$$
A_n(T) = \frac{\sinh \beta_\pm}{\cosh \left[  \beta_\pm (n-x_\pm ) \right]}e^{i
\sigma_\pm}.
$$

We can investigate the linear stability of these solutions.
The linearization of the nonlinear system (\ref{eq:al:betaT}-\ref{eq:al:sigmaT})
around the parameters of the
stationary solitons gives the linear system
\begin{eqnarray}
\label{eq:beta1T}
{\beta_1}_T &=& -2 \Delta \sinh^2 \beta_\pm \psi_{\beta_\pm}(x_\pm) \cos(2 \sigma_\pm) \alpha_1
- 4 \Delta \tanh\beta_\pm  \cos(2 \sigma_\pm) \sigma_1 \, ,\\
{\alpha_1}_T& =& - 2\Gamma \alpha_1 \, ,\\
\label{eq:sigma1T}
 {\sigma_1}_T &=& 2\sinh \beta_\pm \beta_1 -2 \Gamma \sigma_1 \, ,
 \\
 \nonumber
 {x_1}_T &=& 2\frac{\sinh \beta_\pm}{\beta_\pm} \alpha_1 -
\Delta \frac{\sinh^2 \beta_\pm}{\beta_\pm}
\frac{\pi^2+4\beta_\pm^2}{3\beta_\pm^2 \sinh(2 \beta_\pm)} \cos(2 \sigma_\pm) \alpha_1 \\
&& -\Gamma \frac{\sinh^2 \beta_\pm}{\beta_\pm} \psi_{\beta_\pm}'
(x_\pm) x_1 - \Gamma \frac{\beta_\pm \sinh(2\beta_\pm) -\sinh^2
\beta_\pm}{\beta_\pm^2} \psi_{\beta_\pm}(x_\pm) \beta_1  \, .
\end{eqnarray}
The equation for $x_1$ is decoupled from the other ones.
The three eigenvalues for the $3\times3$ system
for $(\alpha_1,\beta_1,\sigma_1)$ are
\begin{eqnarray*}
\lambda_\pm^{(1)} &=& -2 \Gamma \, ,\\
\lambda_\pm^{(2)} &=& - \Gamma +\sqrt{\Gamma^2 -8 \Delta \sinh\beta_\pm \tanh \beta_\pm \cos(2\sigma_\pm)}  \, ,\\
\lambda_\pm^{(3)} &=& -  \Gamma - \sqrt{\Gamma^2-8 \Delta \sinh\beta_\pm \tanh \beta_\pm \cos(2\sigma_\pm)}  \,  .
\end{eqnarray*}
The fixed point is stable if the real parts of the eigenvalues
are nonpositive.
This shows that the fixed point labeled $-$ is not stable, since
$\lambda^{(2)}_- >0$,
while the fixed point labeled $+$ is stable.

In absence of damping $\Gamma=0$ (which is equivalent to $\gamma=0$),
the soliton parameters $\beta$ and $\sigma$ (amplitude and phase)
oscillate with the frequency (in the $T$-variable)
\begin{equation}
\label{eq:periodg0}
\Omega_+ = \sqrt{8 \Delta \sinh\beta_+ \tanh \beta_+}
\end{equation}
while the parameter $\alpha$ (velocity) is constant.
This means that
the stationary soliton is stable.
This also shows that
stable slowly moving breathers can propagate
in the presence of parametric drive.

\begin{figure}
\begin{center}
\begin{tabular}{c}
\includegraphics[width=5.7cm]{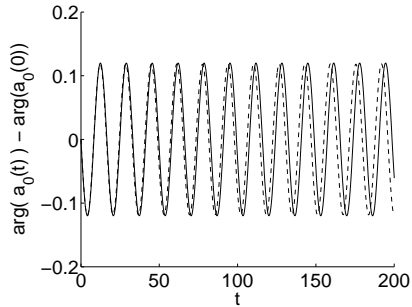}
\end{tabular}
\vspace*{-0.25in}
\end{center}
\caption{Here $\chi=0$, $\omega=1$, $\delta =0.022$, and $\gamma=0$
(i.e. $\Delta=0.0218$ and $\Gamma=0$).
The initial condition is a soliton with $\sigma=\sigma_+$, $x_+=0$,
$\alpha=0$, and
$\beta=\beta_+ +  0.02$ ($\                                                                                                                           \beta_+=0.948$).
We plot the solution phase obtained from the numerical integration
of  Eq.~(\ref{eq:ala})  (dashed line)
and compare with the theoretical oscillation obtained from
 (\ref{eq:beta1T}-\ref{eq:sigma1T})
(solid line).
The theoretical oscillation period (in the $t$-variable)
is $2\pi/\Omega_+ /(1+\delta/2)= 16.6$.
\label{fig1} }
\end{figure}

\begin{figure}
\begin{center}
\begin{tabular}{c}
\includegraphics[width=5.7cm]{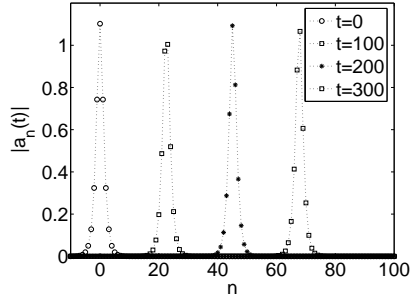}
\end{tabular}
\vspace*{-0.25in}
\end{center}
\caption{Here $\chi=0$,  $\omega=1$, $\delta =0.022$, $\gamma=0$.
The initial condition is a soliton with $\sigma=\sigma_+$, $x_+=0$,
$\alpha=0.1$, and
$\beta=\beta_+ $.
We plot the soliton profiles $|a_n(t)|$ at different times,
which exhibits the stable propagation of the moving soliton.
\label{fig2} }
\end{figure}

We have performed numerical simulations of
Eq.~(\ref{eq:ala}) to confirm these predictions.
In Fig.~\ref{fig1} we consider a perturbation of the initial amplitude.
The periodic oscillations of the soliton parameters have the predicted period
(\ref{eq:periodg0}).
In Fig.~\ref{fig2} we consider a soliton with a positive velocity.
As predicted by the theory, this soliton can propagate in a stable way.

In presence of damping the  soliton parameters $\beta$ and $\sigma$
oscillate with the frequency
\begin{equation}
\label{eq:periodg0b}
\Omega_+ = \sqrt{8\sqrt{\Delta^2 - \Gamma^2} \sinh\beta_+ \tanh
\beta_+- \Gamma^2}.
\end{equation}
These oscillations decay exponentially with the rate
$\Gamma$.
Besides, comparing (\ref{eq:periodg0}) and (\ref{eq:periodg0b})
shows that the damping enhances the oscillation period. The
soliton parameter $\alpha$ decays exponentially at the rate $2
\Gamma$. Therefore, the stationary soliton is very stable.
However, the propagation of moving solitons is not supported, as
the soliton velocity decays exponentially to $0$. If we denote by
$\alpha_0$ the initial value of the parameter $\alpha$, then the
input soliton converges to the stationary form
$$
A_n = \frac{\sinh \beta_+}{\cosh \left[ \beta_+ (n-x_+ -x_F ) \right]}e^{i \sigma_+}
$$
with
\begin{equation}
\label{eq:defxf} x_F  = \frac{\sinh\beta_+ \sin\alpha_0}{\beta_+ \Gamma} .
\end{equation}

\begin{figure}
\begin{center}
\begin{tabular}{c}
\includegraphics[width=5.7cm]{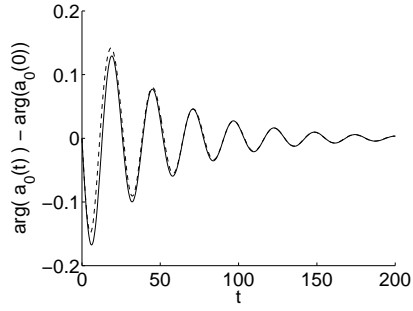}
\end{tabular}
\vspace*{-0.25in}
\end{center}
\caption{Here $\chi=0$, $\omega=1$, $\delta =0.022$, $\gamma=0.02$.
The initial condition is a soliton with $\sigma=\sigma_+$, $x_+=0$,
$\alpha=0$, and
$\beta=\beta_+ +0.02$ ($\beta_+=0.942$).
We plot the solution phase obtained from the numerical integration
of Eq.~(\ref{eq:ala})
(dashed line).
The observed oscillations and damping are correctly predicted by the model
(\ref{eq:beta1T}-\ref{eq:sigma1T}) (solid line).
The period is $2\pi/\Omega_+/(1+\delta/2)=25.9$ and the exponential decay rate
is $\gamma=0.02$
 (solid line).
\label{fig3} }
\end{figure}

\begin{figure}
\begin{center}
\begin{tabular}{c}
\includegraphics[width=5.7cm]{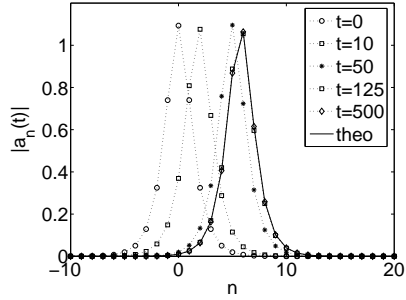}
\end{tabular}
\vspace*{-0.25in}
\end{center}
\caption{Here $\chi=0$, $\omega=1$, $\delta =0.022$, $\gamma=0.02$.
The initial condition is a soliton with $\sigma=\sigma_+$, $x_+=0$,
$\alpha=0.1$, and
$\beta=\beta_+ $.
We plot the soliton profiles $|a_n(t)|$ at different times,
which exhibits the trapping of the moving soliton.
The solid line is the theoretical stable stationary soliton centered at $x_F = 5.76$
given by (\ref{eq:defxf}).
\label{fig4} }
\end{figure}

In Fig.~\ref{fig3}, we
plot the damping of the oscillations of the soliton
parameters and compare it to the theoretical formula.
In Fig.~\ref{fig4} the trapping of a soliton with an initial
velocity is shown. The final position of the soliton
is given by (\ref{eq:defxf}).

\begin{figure}
\begin{center}
\begin{tabular}{c}
\includegraphics[width=5.7cm]{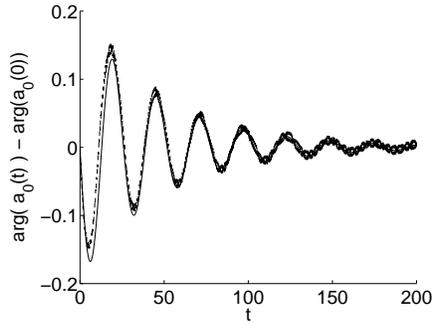}
\end{tabular}
\vspace*{-0.25in}
\end{center}
\caption{Numerical simulation of Eq.~(\ref{eq:al0}). Here
$\chi=0$, $\omega=1$, $\delta =0.022$, $\gamma=0.02$, $f(\tau) =\sin(2 \pi \tau)$
and $\eps=0.125$. The initial condition is a soliton with
$\sigma=\sigma_+$, $x_+=0$, $\alpha=0$, and $\beta=\beta_+ +0.02$.
We plot the phase of $a_0(t):=u_0(t)+i \sin(16 \pi t)$ (dashed line).
The observed oscillations and damping are correctly predicted by
the model, and correspond exactly to the ones of the solution of
the averaged equation (see Fig.~\ref{fig3}). \label{fig3b} }
\end{figure}

We have also simulated the original equation (\ref{eq:al1}) with
the external drive $f(\tau)=\sin(2 \pi \tau)$ and $\eps=0.125$. We plot
one of the obtained results in Fig.~\ref{fig3b} (to be compared
with Fig.~\ref{fig3}), which shows full agreement.

\subsection{Periodic solutions of the damped AL
equation with parametric drive}
In the following we discuss exact
periodic solutions of the parametrically damped and driven AL
equation
\begin{equation} i {{A}_n}_T +
[A_{n+1}- (2+\Omega) A_n + A_{n-1}]+ [i \eta + (1 + i \eta)|A_n|^2]
(A_{n+1}+A_{n-1})  = -i \Gamma A_n + \Delta \overline{A_n}
, \label{EQ:SAL4}
\end{equation}
where $\eta \geq 0$ models a small dispersive and nonlinear damping.
\begin{figure}[htb]
\centerline{
\includegraphics[width=4.cm,height=4.cm,clip]{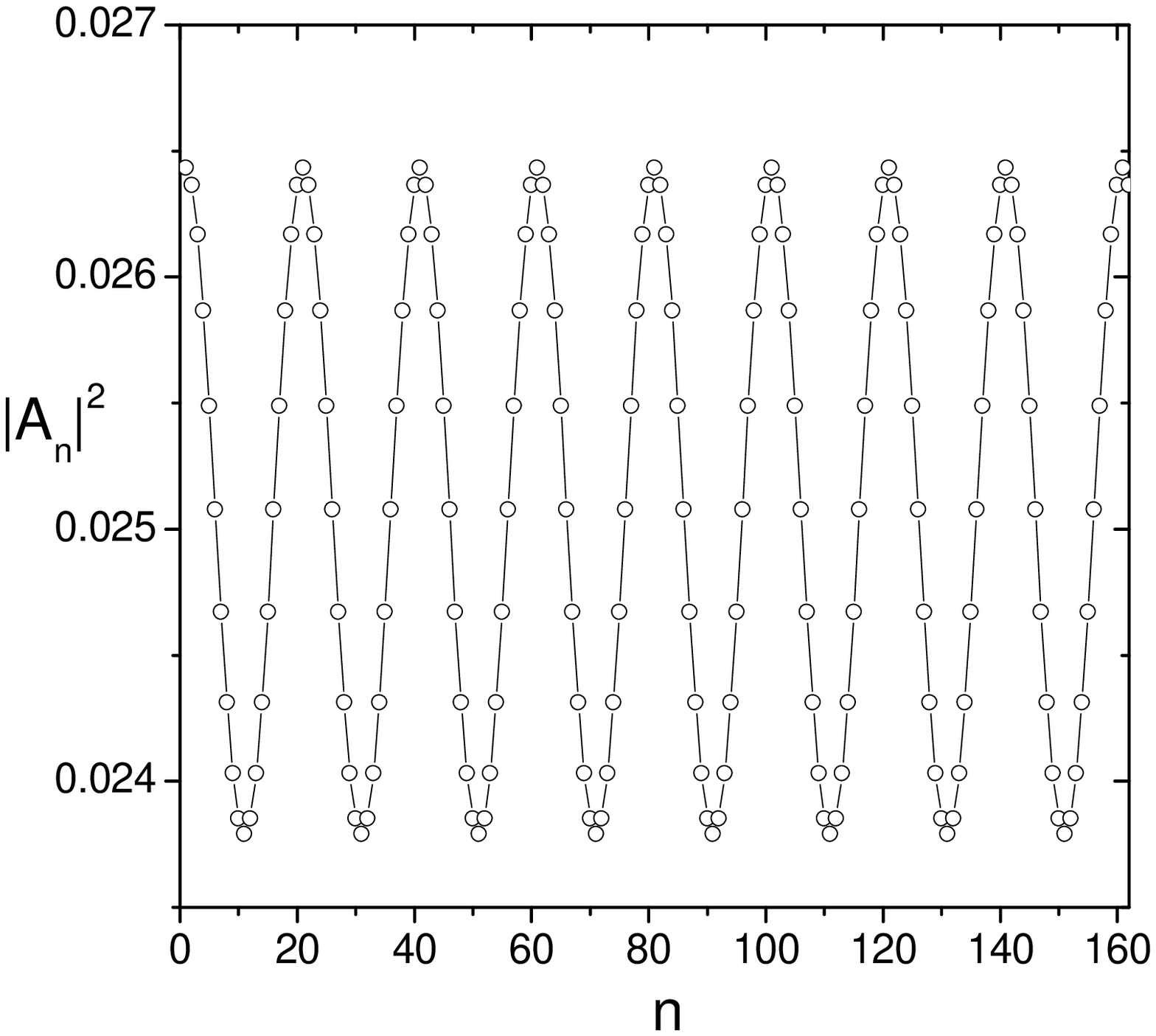}
\includegraphics[width=4.cm,height=4.cm,clip]{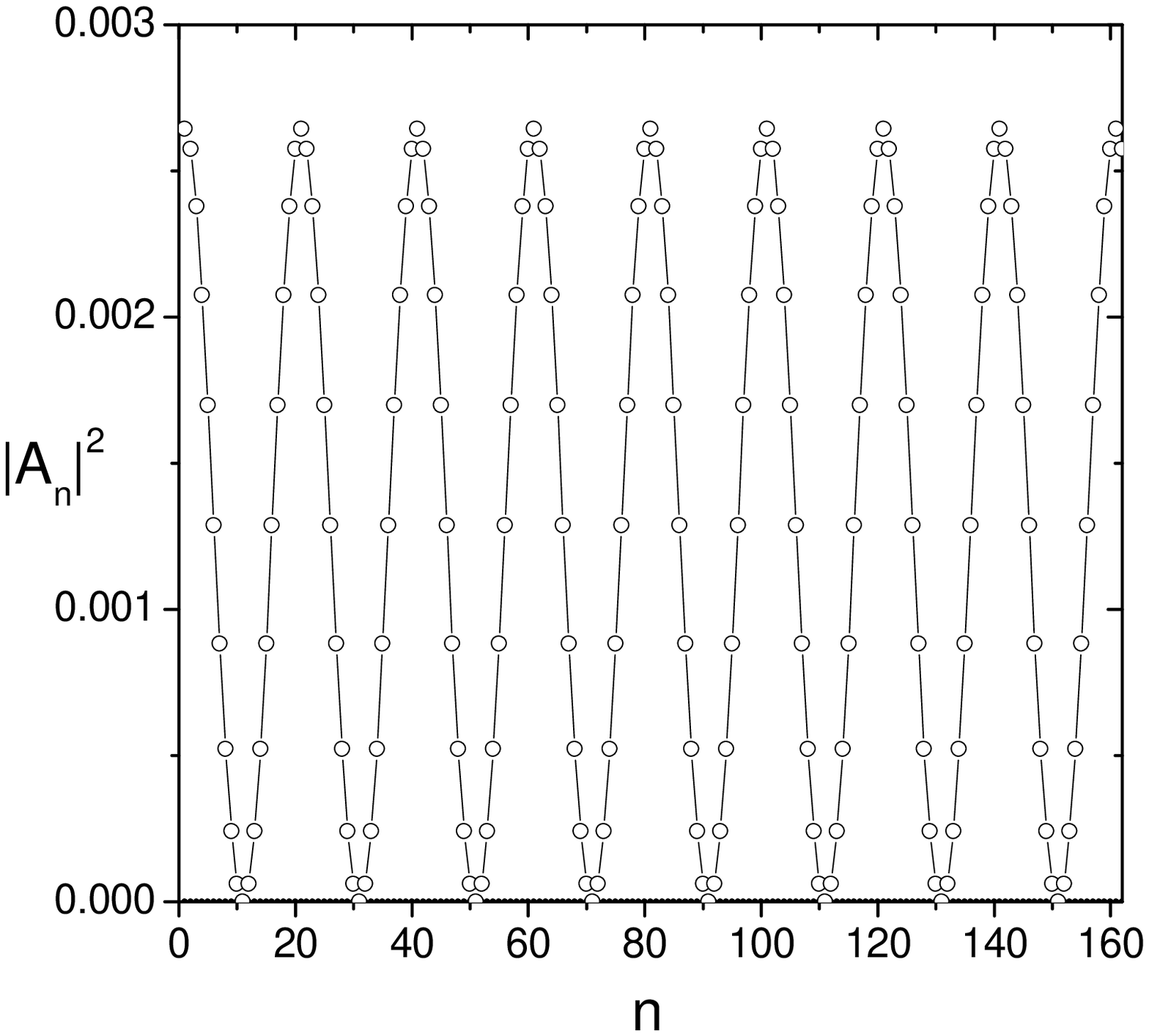}
} \caption{Left panel: Square modulus of the $\dn$ solution
in Eq.~(\ref{SOL:DN}) with
 $m=0.1$, $\beta=0.161244$ ($N_p=20$), $x_F=0$,
 on a line of $160$
points. The parameters are fixed as
$\Delta = 0.04 $, $\Gamma=0.02$, $\Omega=0.01558$, $\eta=0$.
Right panel: Square modulus of the $\cn$
solution in Eq.~(\ref{SOL:CN}) with parameters fixed as in the left
panel.} \label{figms1}
\end{figure}
To this regard we consider periodic stationary solutions  of the
form
 \be A_n = A
e^{i \sigma }\dn\left[ \beta(n+x_F),m \right]\,. \label{SOL:DN}
\ee
Direct substitution of Eq. (\ref{SOL:DN}) into Eq.
(\ref{EQ:SAL4}) shows that a solution is indeed obtained provided
the two following relations are satisfied
by the soliton parameter $\beta$ and the modulus $m$:
\begin{eqnarray}
\label{SOL:par-a}
&& 2 + \Omega
- 2 \frac { \dn(\beta,m)}{\cn(\beta,m)^2} + \Delta
\left[ 1- \left(\frac{\Gamma}{\Delta}  + \frac{2\,\eta\,\dn(\beta
,m)}{\Delta\,\cn(\beta ,m)^2}\right)^2 \right]^{1/2} = 0 , \\
&& \beta N_p= 2 K(m) ,
\label{SOL:par-b}
\end{eqnarray}
where $N_p$ is the number of sites in
one period  and $K(m)$ is
the complete elliptic integral of the first kind.
Here $m \in (0,1)$ and $N_p$ must be a positive integer,
so that there exists a numerable set of pairs $(\beta,m)$ that satisfy
the conditions (\ref{SOL:par-a}-\ref{SOL:par-b}).
If the conditions are fulfilled, then the periodic function (\ref{SOL:DN}) is a solution
of (\ref{EQ:SAL4}) with the amplitude $A$ and phase $\sigma$
given by
\be
A =\pm
\frac{\sn(\beta,m)}{\cn(\beta,m)},\;\;\; \sigma=-\frac12
\arcsin\left(\frac{\Gamma}{\Delta}  + \frac{2\,\eta\,\dn(\beta
,m)}{\Delta\,\cn(\beta ,m)^2}\right) ,
 \label{SOL:par2}
 \ee
while the soliton center $x_F$ is arbitrary.

Another periodic solution of  Eq. (\ref{EQ:SAL4}) can be
constructed by replacing the function $\dn$ into the ansatz
(\ref{SOL:DN}) with the elliptic cosine $\cn$:
\be
A_n = A  e^{i \sigma } \cn \left[ \beta(n+x_F),m \right]\,.
\label{SOL:CN}
\ee
 In this case
the two conditions to be satisfied by $\beta$ and $m$ are
\begin{eqnarray}
&& 2 + \Omega
- 2 \frac { \cn(\beta,m)}{\dn(\beta,m)^2} + \Delta
\left[ 1 - \left( \frac{\Gamma}{\Delta}  +  \frac{ 2 \eta \cn(\beta
,m)}{\Delta \dn(\beta ,m)^2}\right)^2 \right]^{1/2}
=0 , \\
&& \beta N_p= 4 K(m) ,
\end{eqnarray}
and the soliton amplitude and phase  are
\be
A=\pm \sqrt{m}
\frac{\sn(\beta,m)}{\dn(\beta,m)},\;\;\; \sigma=-\frac12
\arcsin\left(\frac{\Gamma}{\Delta}  + \frac{2\,\eta\,\cn(\beta
,m)}{\Delta\,\dn(\beta ,m)^2}\right).
 \ee
\begin{figure}[htb]
\centerline{
\includegraphics[width=4.cm,height=4.cm,clip]{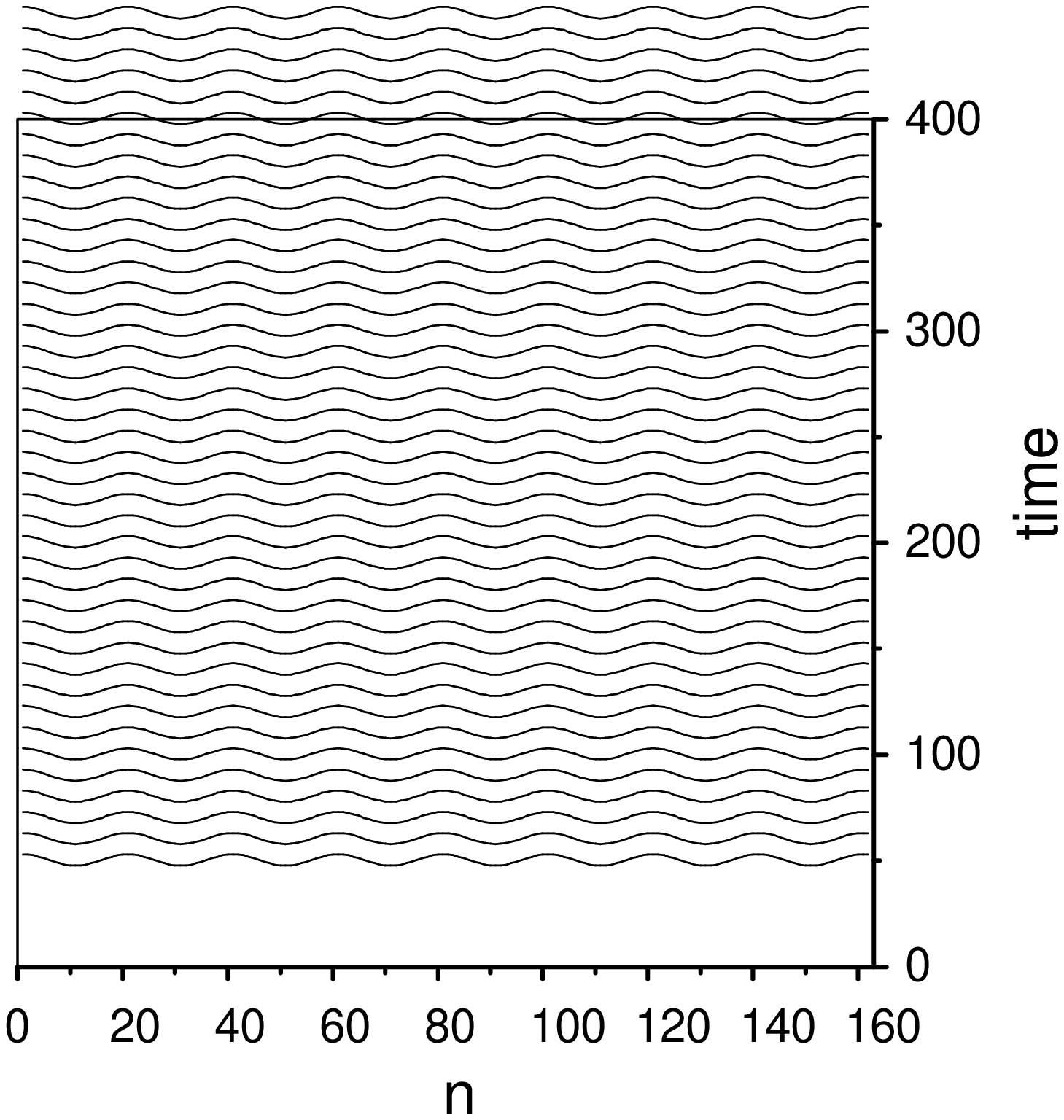}
\includegraphics[width=4.cm,height=4.cm,clip]{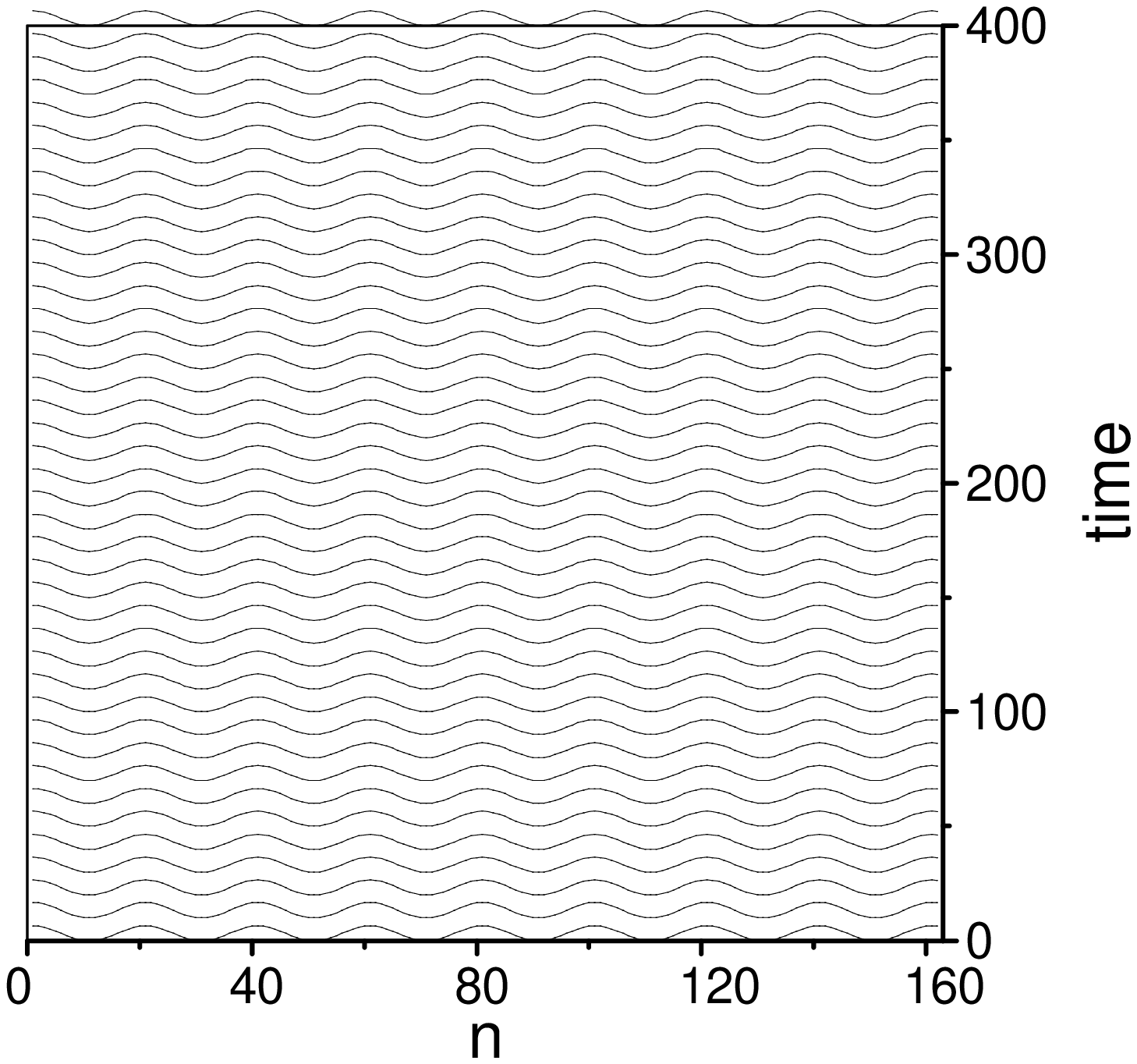}
} \caption{Time evolution of $|A_n|^2$ for of the $\dn$ (left
panel) and $\cn$ (right panel) solutions in Fig.~\ref{figms1}.}
\label{figms2}
\end{figure}

It may be worth noting that in deriving these solutions usage of
the following identities for the Jacobi elliptic functions
\cite{kare} have been made:
\bea
\dn(x+a,m)+\dn(x-a,m) &=& \frac{2
\dn(a,m)\, \dn(x,m)}{\cn(a,m)^2+ \sn(a,m)^2 \dn(x,m)^2}, \\
\cn(x+a,m) + \cn(x-a,m) &=& \frac{2\,\cn(a,m) \,
\cn(x,m)}{\dn(a,m)^2 + m\, \sn(a,m)^2\, \cn(x,m)^2}.
\eea
Also
notice that in the limit of infinite period (i.e. $m \rightarrow
1$) the above solutions both reduce to the AL soliton
\be
A_n=\frac{\sinh(\beta)}{\cosh \left[ \beta (n+x_F) \right]}e^{i \sigma}
\label{ALsoliton}
\ee
with the soliton parameter satisfying
$$
2+\Omega+
\left[ \Delta^2-\left(\Gamma  +
2\eta\cosh(\beta) \right)^2 \right]^{1/2}- 2 \, \cosh(\beta) =0.
$$
Similar periodic solutions exist also for the unperturbed AL
equation \cite{sharf} and for generalized AL equations with
arbitrarily high-order nonlinearities \cite{krsss}.

In Fig. \ref{figms1} we depict the waveforms of the above
solutions on a line of $160$ points for the case
$\Delta = 0.04 $, $\Gamma=0.02$, $\Omega=0.01558$, $\eta=0$,
and the solution parameters are $m=0.1$, $\beta=2
K(m)/N_p=0.161244$, $N_p=20$.
\begin{figure}[htb]
\centerline{
\includegraphics[width=4.cm,height=4.cm,clip]{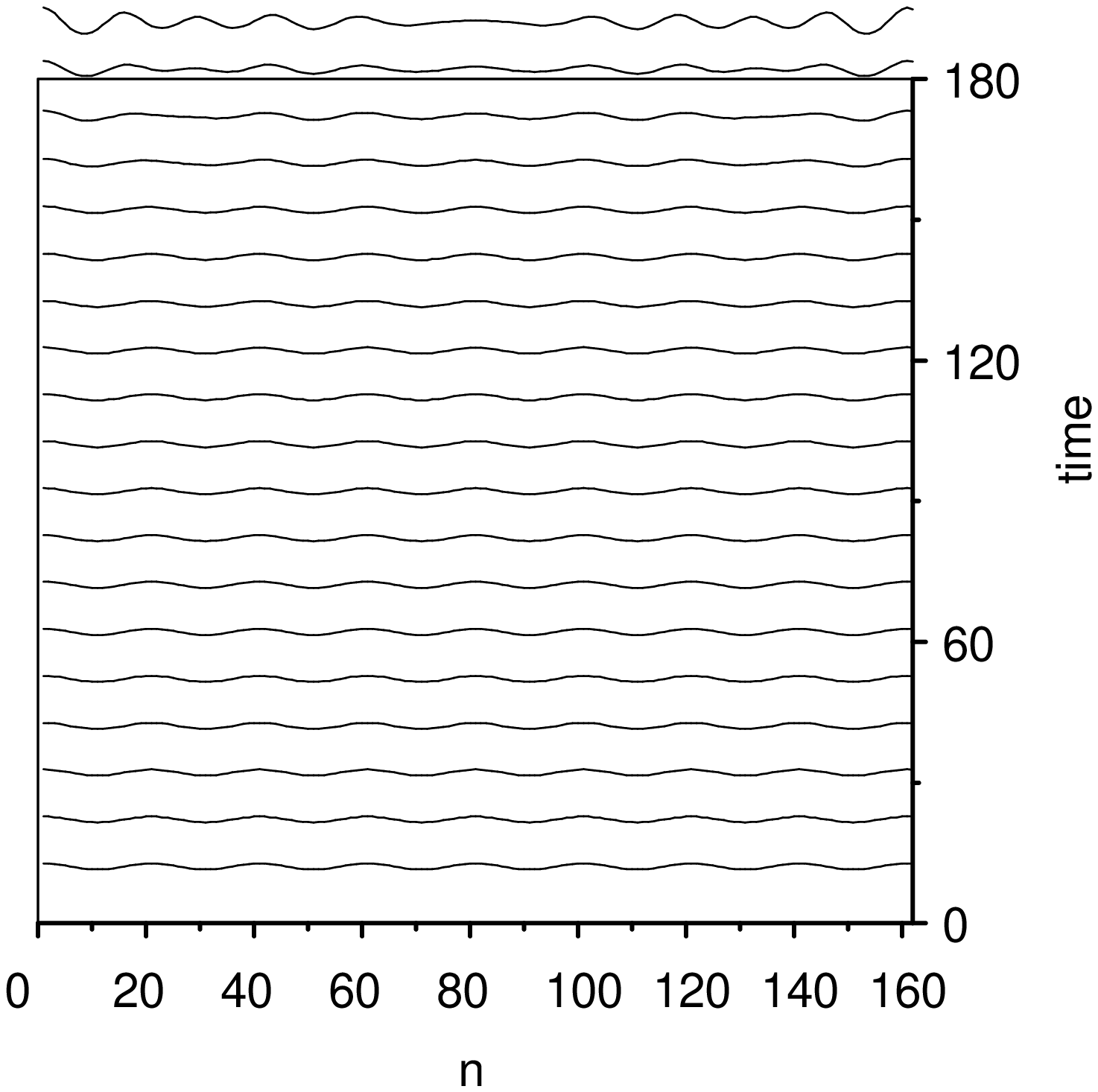}
\includegraphics[width=4.5cm,height=4.5cm,clip]{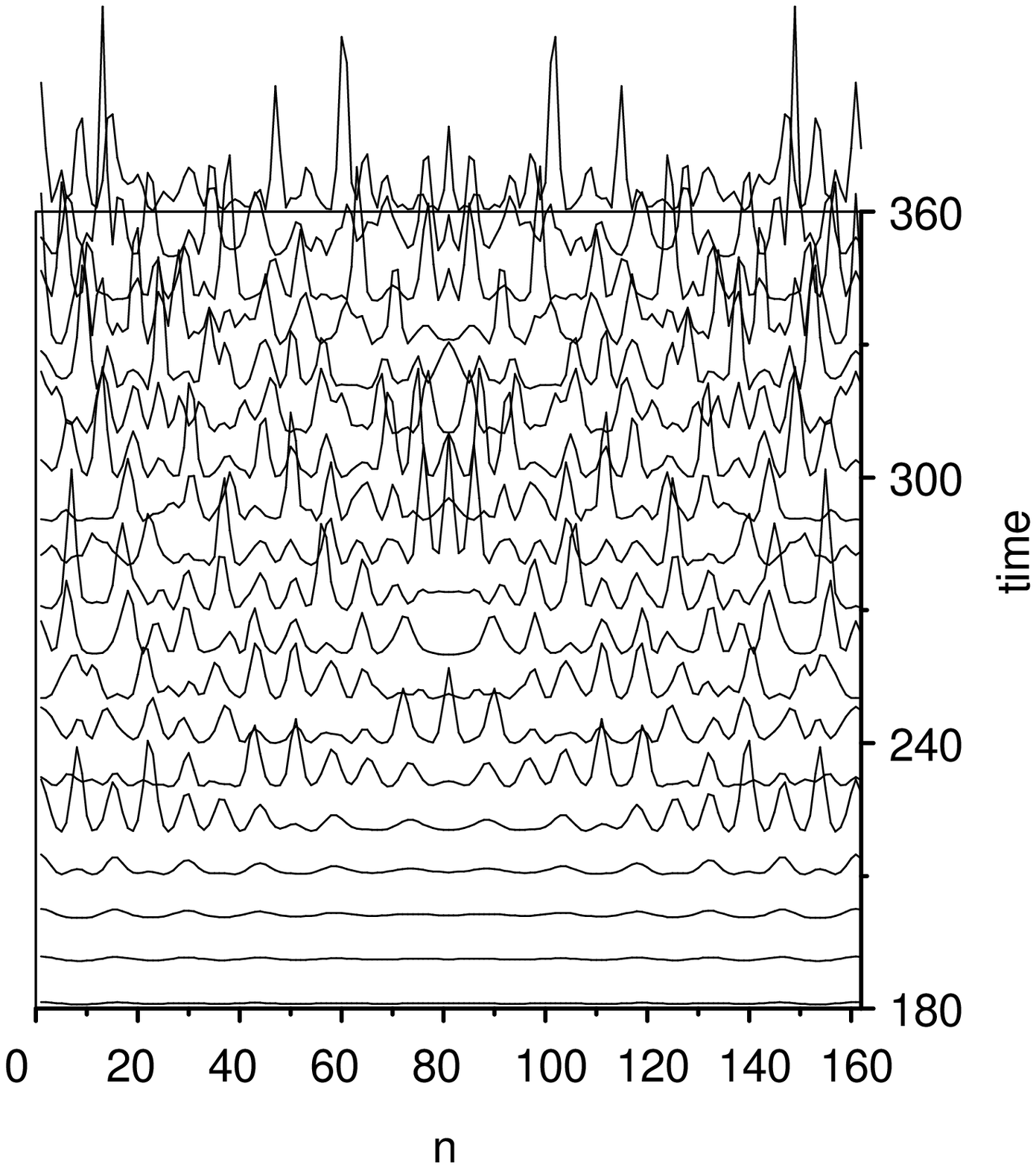}}
\caption{Time evolution of the unstable  dn solution in Eq.~(\ref{SOL:DN})
obtained for
$\Delta = 0.15$, $\Gamma=0.02$, $\Omega=-0.098438$, $\eta=0$,
and the solution parameters are as in Fig.~\ref{figms1}:
$m=0.1$, $\beta=2
K(m)/N_p=0.161244$, $N_p=20$.
 Notice the change in the
$|A_n|^2$ scale from one panel to another. }
\label{figms3}
\end{figure}
We find, by direct numerical integrations of Eq.~(\ref{EQ:SAL4}),
that for small values of the damping and parametric driver
amplitude these solutions remain stable under very long time
evolution. This is shown in Fig.~\ref{figms2} where the time
evolution obtained from direct numerical simulations of
Eq.~(\ref{EQ:SAL4}) is depicted for parameters
$\Delta = 0.15$, $\Gamma=0.02$, $\Omega=-0.098438$, $\eta=0$,
which supports the same dn and cn solutions as in Fig.~\ref{figms1}.
By keeping fixed the damping
constant and increasing the amplitude of the parametric driver in
a certain range, we find that  these solutions remain
stable for very long time, while outside of this range
instabilities quickly develop. The development of the instability
for out of range parameter is investigated in Figs.~\ref{figms3}-\ref{figms4}.
\begin{figure}[htb]
\centerline{
\includegraphics[width=4.cm,height=4.cm,clip]{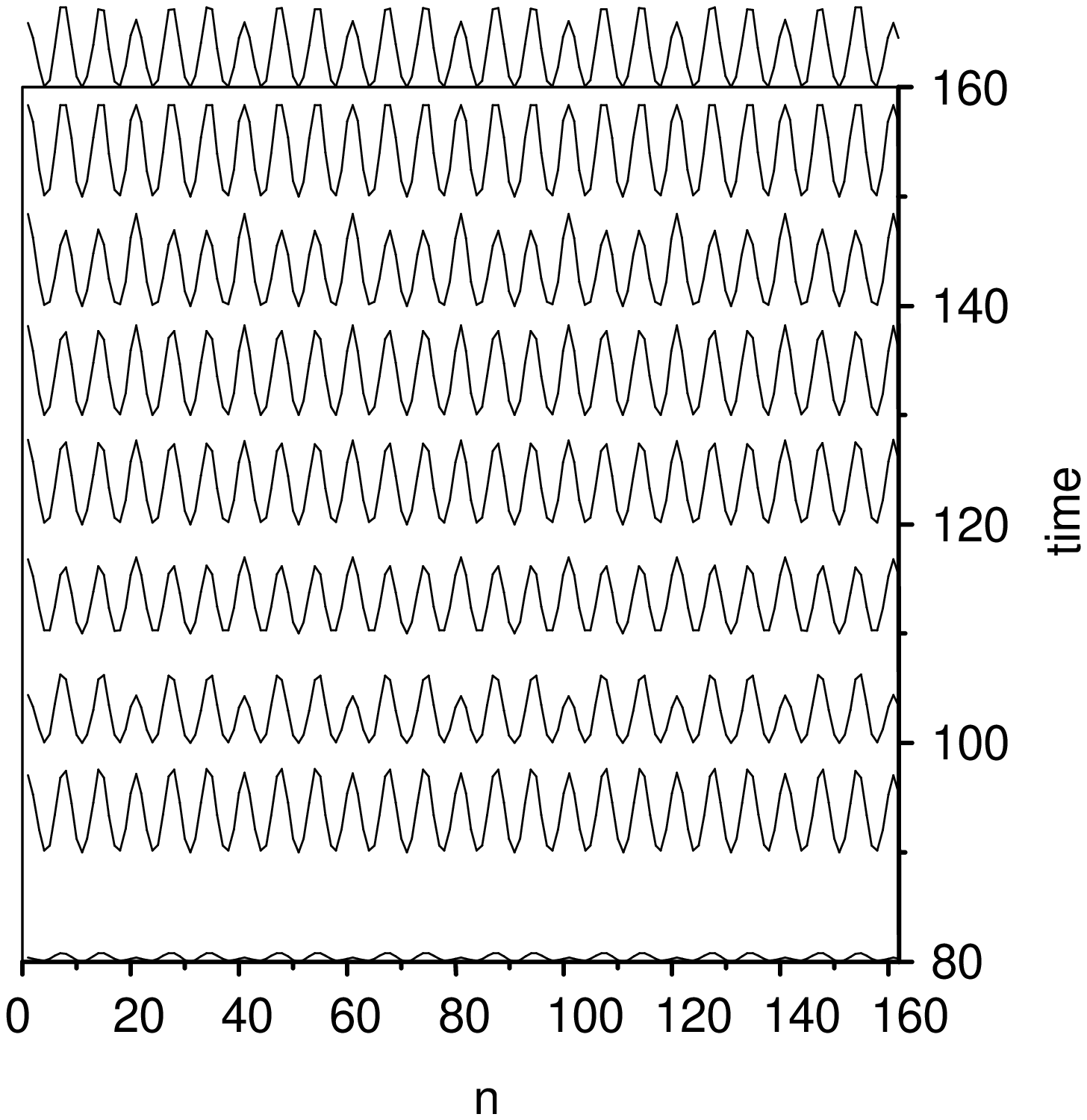}
\includegraphics[width=4.cm,height=4.cm,clip]{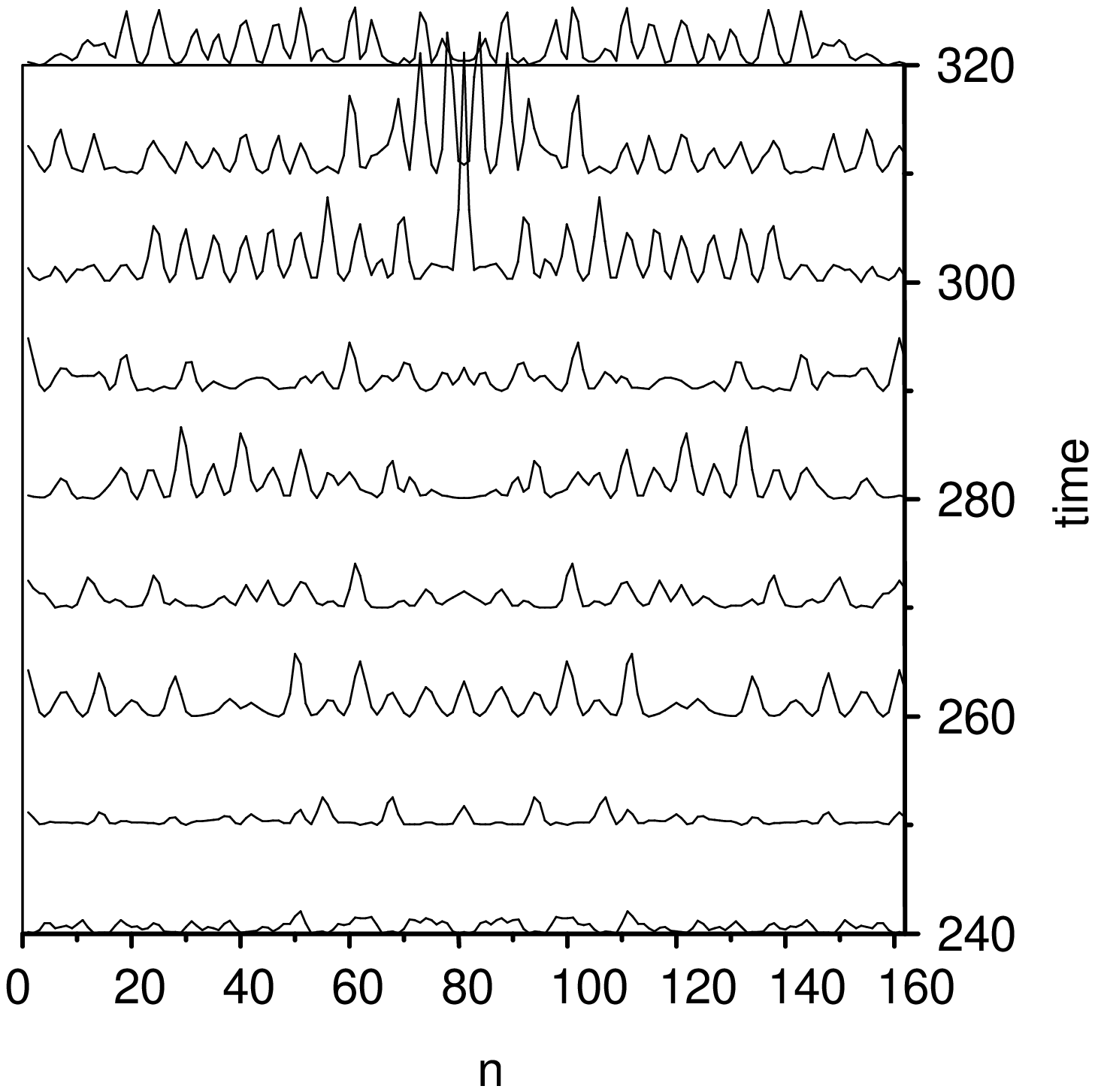}}
\centerline{
\includegraphics[width=4.cm,height=4.cm,clip]{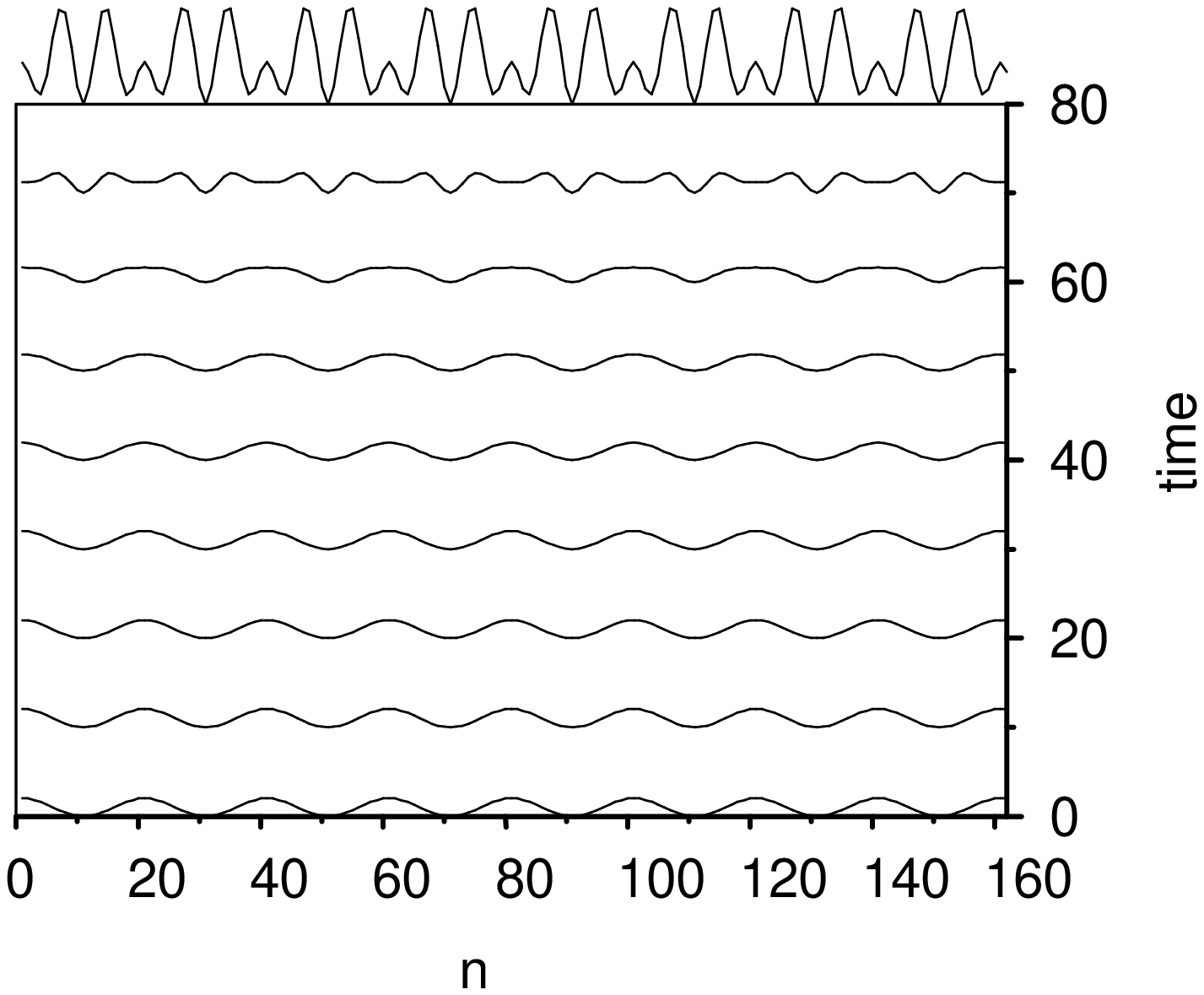}
\includegraphics[width=4.cm,height=4.cm,clip]{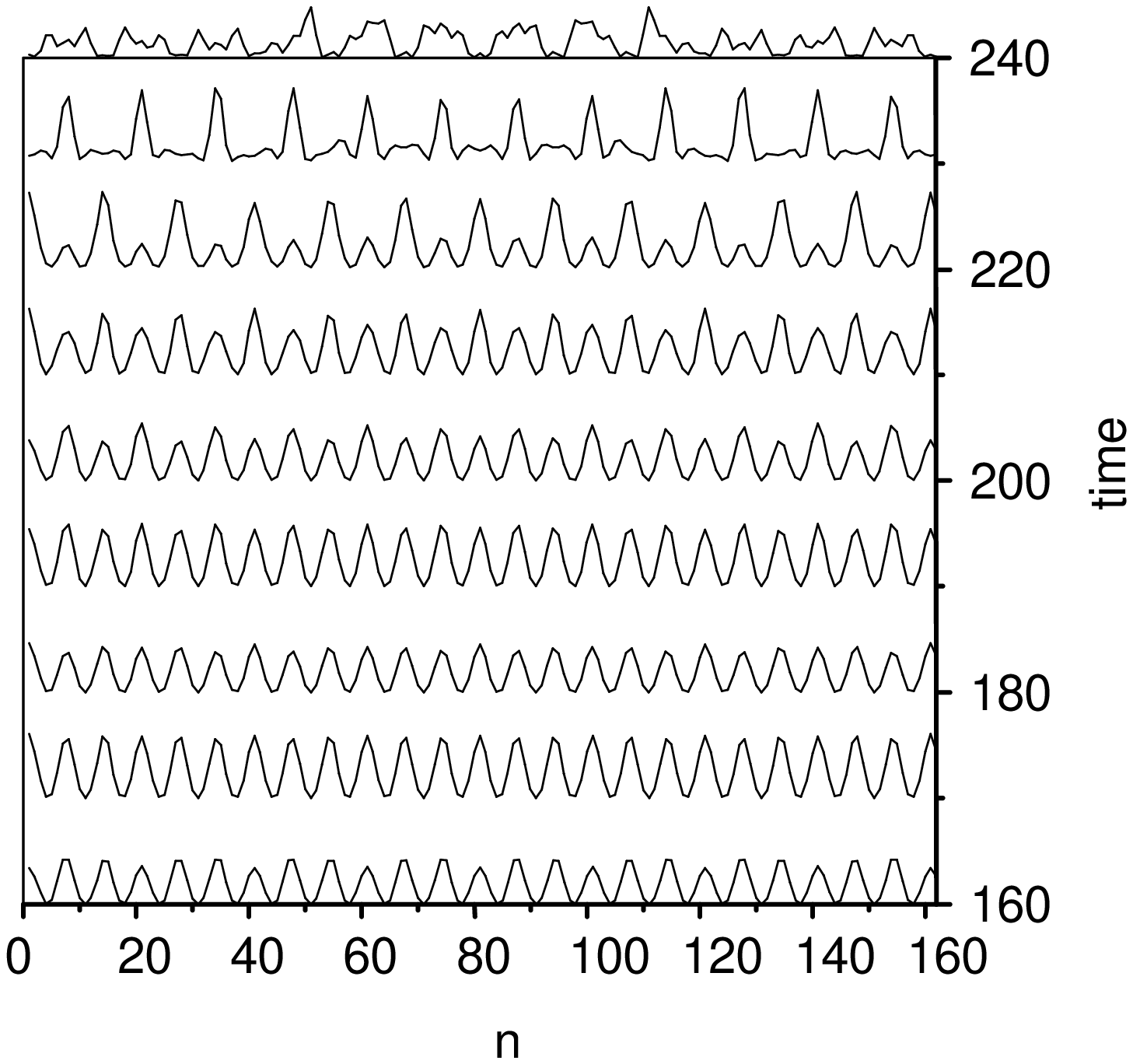}}
\caption{Time evolution of the unstable  cn solution in Eq.~(\ref{SOL:CN})
when the parameters are
fixed as in Fig.~\ref{figms1}.
Notice the change in the
$|A_n|^2$ scale passing from one panel to another. \label{figms4}
}
\end{figure}

\begin{figure}[htb]
\centerline{
\includegraphics[width=4.cm,height=4.cm,clip]{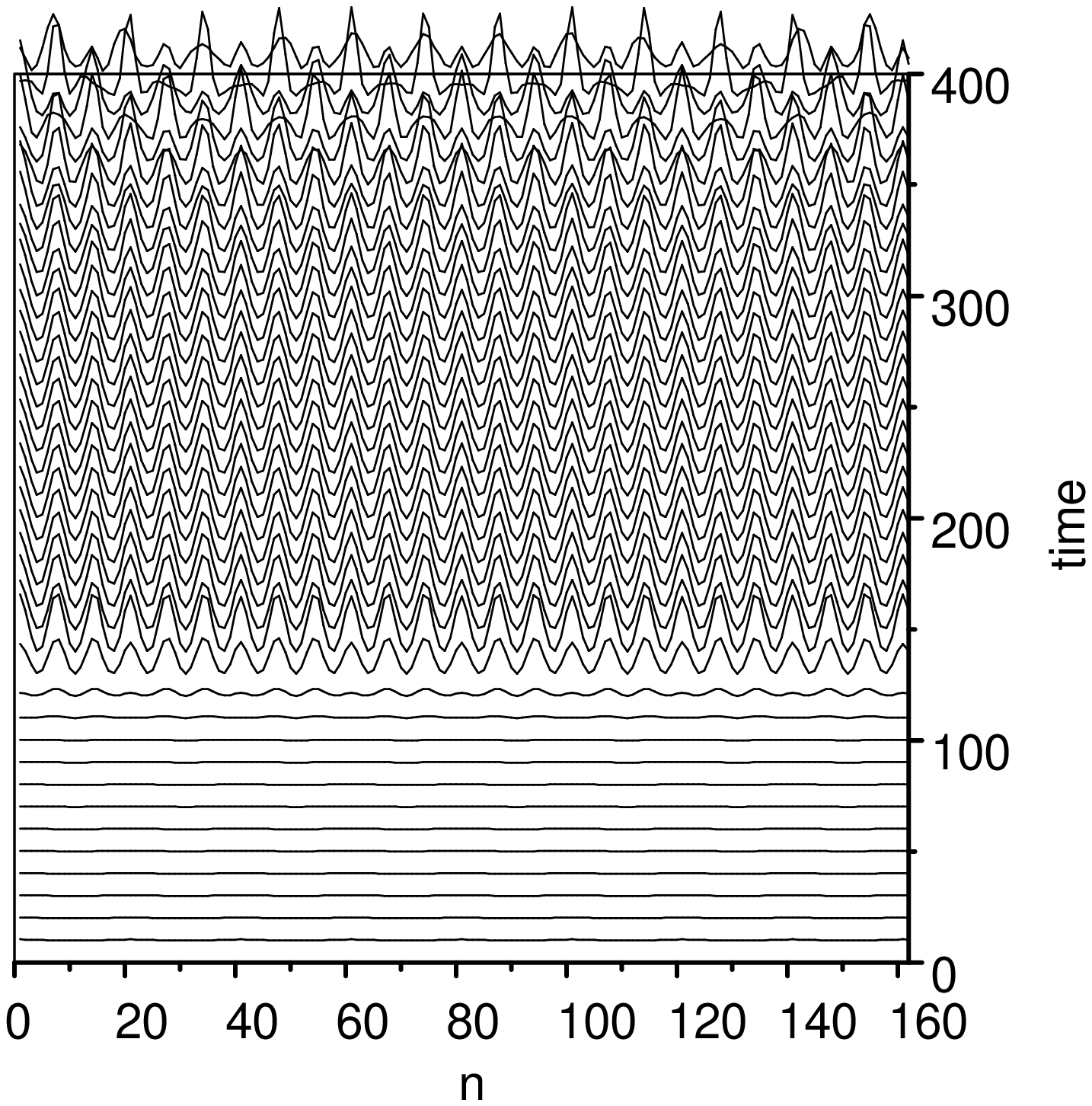}}
\caption{Time evolution of the unstable  dn solution in Eq.~(\ref{SOL:DN})
obtained for $\eta=0.02$. Other parameters are fixed
as in Fig.~\ref{figms4}.} \label{figms5}
\end{figure}

Notice that while for the dn solution the instability suddenly
sets in without any apparent pattern, the instability of the cn
solution seems to follow a precise pattern. In particular, from
Fig. \ref{figms4} we see  that before the instability fully
develops at time $t\approx 200$ the cn solution bifurcates into a
period three solution at $t\approx 80$ which remains stable for a
long time. The presence of a small dispersive
nonlinear damping (controlled by the parameter
$\eta$) effectively increases the stability of both the period one
and  the period three solutions, as one can see from
Fig.~\ref{figms5}. The scenario behind the development of the
instability of the cn solution is quite interesting and deserves
more investigations.

\section{The damped DNLS system with parametric drive}
We consider in this section that the nonlinear term is not of the
AL type but it is a mixture of the AL cubic intersite nonlinearity
and the onsite DNLS cubic term. We shall analyze this system by
considering such a nonlinearity as a perturbation of the cubic AL
nonlinearity.

\subsection{Perturbation theory for the cubic nonlinearity}
In this subsection
we consider the perturbed AL equation (\ref{eq:al1})
when the perturbation is given by
\begin{equation}
\label{eq:pert2}
R_n = \chi |A_n|^2 ( A_{n+1} +A_{n-1}  -  2A_n )
\end{equation}
If $\chi =0$, then Eq.~(\ref{eq:al1}) with the perturbation
(\ref{eq:pert2}) is the integrable
AL equation.
If $\chi =1$, then it is the standard DNLS equation.
In the adiabatic approximation, this perturbation has no effect
on the first-order differential equation for the
amplitude soliton parameter $\beta$,
but the evolution equations for the parameters $x$, $\alpha$, and $\sigma$
have corrective terms
\begin{eqnarray}
x_T &=& 2(1-\chi)  \frac{\sinh \beta}{\beta }\sin \alpha +2 \chi \frac{\sinh^2 \beta}{\beta^2 \cosh \beta} \sin \alpha \\
\alpha_T &=& \chi P_\beta(x)
 \\
\sigma_T &=& 2\cosh \beta \cos \alpha
+2 \chi (1-\cos \alpha) \sinh \beta \tanh \beta
\nonumber\\
&&+2 (1-\chi)
\alpha \sin \alpha \frac{\sinh \beta}{\beta}
 +2\chi\alpha \sin \alpha \frac{\sinh^2 \beta}{\beta^2  \cosh\beta }  - 2- \Omega
+ 2 \chi Q_\beta(x)
\end{eqnarray}
where
\begin{eqnarray}
\label{eq:Pbeta}
P_\beta(x) &=&
  \sum_{s=1}^\infty
\frac{ 8 \pi^3 s^2 \sinh^2 \beta}
{\beta^3 \sinh (\frac{\pi^2 s }{\beta})}
\sin (2 \pi s x)\\
\nonumber
Q_\beta(x)  &=&  -1 +\frac{\sinh 2\beta}{\beta} - \frac{\sinh^2 \beta}{\beta^2}
-
 \sinh \beta \tanh \beta \\
&& + 2 \sum_{s=1}^\infty \frac{ 2 \pi^2 s \beta^2 \sinh \beta
\cosh \beta +[  \pi^4 s^2 {\rm cotanh}(\frac{\pi^2 s}{\beta}) -2
\pi^2 s \beta] \sinh^2 \beta} {\beta^4 \sinh (\frac{\pi^2 s
}{\beta})} \cos (2 \pi s x). \label{eq:Qbeta}
\end{eqnarray}
We should keep in mind that the adiabatic approximation is valid when the
perturbation (\ref{eq:pert2}) is small,
which is true if $\chi$ is small and $\beta$ is arbitrary, or if
 $\chi$ is arbitrary and $\beta$ is small.
 In the following, we shall only keep the term $s=1$ in the sums
 (\ref{eq:Pbeta}-\ref{eq:Qbeta}),
to simplify the algebra, although the analysis could be carried out
with the full expressions.
This simplification is consistent with the adiabatic approximation.

\subsection{Parametrically driven DNLS solitons}
We now consider that the perturbation $R_n$ is given by (\ref{eq:pert0}),
that is  the sum of the
cubic perturbation (\ref{eq:pert2}) and the parametric drive with damping (\ref{eq:pert1}).
In these conditions,Ê
there are two fixed points if $\Gamma <\Delta$
(which is equivalent to $\gamma<\delta$)
and $\Omega+\Delta>0$ (which is equivalent to $\omega-\delta>0$):
\begin{eqnarray}
&&\alpha_{\pm}=0\, ,\ \ \ \ \sin (2\sigma_{\pm}) =-\frac{\Gamma}{\Delta}\, , \ \ \ \
\cos(2 \sigma_{\pm} )= \pm \sqrt{1-\frac{\Gamma^2}{\Delta^2}}\, , \\
&& \cosh(\beta_\pm ) -1 +\chi Q_{\beta_\pm}(0)
=  \frac{\Omega}{2} +\frac{1}{2}
 \sqrt{\Delta^2 -\Gamma^2} \, .
 \label{eq:defbetapm2}
\end{eqnarray}
The center of the soliton $x_\pm$ must be an integer
as soon as $\chi>0$. This is a manifestation of the Peierls-Nabarro
 barrier \cite{vakh1,vakh2}.
Note that the function $\beta \mapsto \cosh(\beta ) -1 +
\chi Q_{\beta}(0)$ is a one-to-one increasing
function from $(0,\infty)$ to $(0,\infty)$
for any $\chi \geq 0$.
Therefore, the parameter $\beta_\pm$ is uniquely determined.
The picture is the same as in the perturbed AL case.
The only difference is a renormalization of the amplitude parameter $\beta_\pm$
given by (\ref{eq:defbetapm2}) instead of (\ref{eq:defbetapm1}).

We next perform the linear stability analysis of the fixed points.
The linearization of the system of ordinary differential equations
for the soliton parameters around the stationary points gives:
\begin{eqnarray}
{\beta_1}_T &=& - 4 \Delta \tanh\beta_\pm  \cos(2 \sigma_\pm) \sigma_1 \, ,\\
{\alpha_1}_T& =& - 2\Gamma \alpha_1  + 16 \pi^4 \chi \frac{\sinh^2 \beta_\pm}{\beta_\pm^3 \sinh( \frac{\pi^2}{\beta_\pm})}  x_1 \, ,\\
 {\sigma_1}_T &=& 2
 \left[ \sinh \beta +\chi
  \partial_\beta Q_\beta(0)\right]_{\beta=\beta_\pm}
  \beta_1
 -2 \Gamma \sigma_1 \, ,
 \\
 \nonumber
 {x_1}_T &=& 2(1-\chi) \frac{\sinh \beta_\pm}{\beta_\pm} \alpha_1
 +2 \chi \frac{\sinh^2 \beta_\pm}{\beta_\pm^2\cosh \beta_\pm} \alpha_1 -
\Delta \frac{\tanh \beta_\pm}{\beta_\pm^3}
\frac{\pi^2+4\beta_\pm^2}{6} \cos(2 \sigma_\pm) \alpha_1
\\
&& -\Gamma \frac{\tanh \beta_\pm}{\beta_\pm^2}
\frac{4\pi^2}{\sinh (\frac{\pi^2}{\beta_\pm})}
 x_1.
 \end{eqnarray}
This $4\times 4$ linear system can be decomposed into two $2\times 2$
linear systems,
for $(\beta_1,\sigma_1$) and for $(\alpha_1,x_1)$, respectively.
It is then easy to compute the eigenvalues. They are:
\begin{eqnarray*}
\lambda_\pm^{(1)}&=&  - \Gamma + \Omega_\pm
\, , \\
\lambda_\pm^{(2)} &=& -  \Gamma - \Omega_\pm  \,  , \\
\lambda_\pm^{(3)}&=&  - \Gamma \left(1 +  \frac{\tanh \beta_\pm}{\beta_\pm^2}
\frac{ 2 \pi^2}{\sinh(\frac{\pi^2}{\beta_\pm})} \right) + \tilde{\Omega}_\pm
\, , \\
\lambda_\pm^{(4)} &=& -  \Gamma \left(1 +\frac{\tanh
\beta_\pm}{\beta_\pm^2} \frac{2
\pi^2}{\sinh(\frac{\pi^2}{\beta_\pm})} \right) -
\tilde{\Omega}_\pm \, ,
\end{eqnarray*}
where the complex numbers $\Omega_\pm$ and $\tilde{\Omega}_\pm$ are given by
\begin{eqnarray}
\label{eq:dnls:omega+}
\Omega_\pm^2 &=&
\Gamma^2 -8 \Delta
 \tanh \beta_\pm \cos(2\sigma_\pm)
[\sinh \beta +
 \chi \partial_\beta Q_\beta(0)]_{\beta=\beta_\pm} \, , \\
 \nonumber
 \tilde{\Omega}_\pm^2 &=& \Gamma^2 \left(1 - \frac{\tanh \beta_\pm}{\beta_\pm^2}
 \frac{2 \pi^2}{\sinh(\frac{\pi^2}{\beta_\pm})} \right)^2 \\
 &&+16 \pi^4
\chi \frac{\sinh^2 \beta_\pm}{\beta_\pm^3 \sinh(
\frac{\pi^2}{\beta_\pm})}
 \left( 2 (1-\chi)
 \frac{\sinh \beta_\pm}{\beta_\pm}
 +2\chi
\frac{\sinh^2 \beta_\pm}{\beta_\pm^2\cosh \beta_\pm} \right.
\nonumber
\\
&& \hspace*{1.55in} \left.
 -\Delta \frac{\tanh \beta_\pm}{\beta_\pm^3}
\frac{\pi^2+4\beta_\pm^2}{6} \cos(2 \sigma_\pm) \right)
 \end{eqnarray}

 The eigenvalues $\lambda_\pm^{(1)}$ and $\lambda_\pm^{(2)}$ describe the growth
 rates of the perturbations of the amplitude parameter $\beta$ and phase parameter $\sigma$.
 The eigenvalues $\lambda_\pm^{(3)}$ and $\lambda_\pm^{(4)}$ describe the growth
 rates of the perturbations of the velocity parameter $\alpha$
 and soliton center $x$.

The function $\beta \mapsto \sinh(\beta ) + \chi \partial_\beta
Q_{\beta}(0)$ is positive valued. Therefore the real part of the
eigenvalue $\lambda_-^{(1)}$ is positive and the stationary point
labelled $-$ is unstable. Besides, the real parts of the
eigenvalues $\lambda_+^{(1)}$ and $\lambda_+^{(2)}$ are
non-positive for any $\Gamma \geq 0$. The real parts of
$\lambda_+^{(3)}$ and $\lambda_+^{(4)}$ are also non-positive,
which implies that
 the stationary point labelled $+$ is stable.

\begin{figure}
\begin{center}
\begin{tabular}{c}
\includegraphics[width=5.0cm]{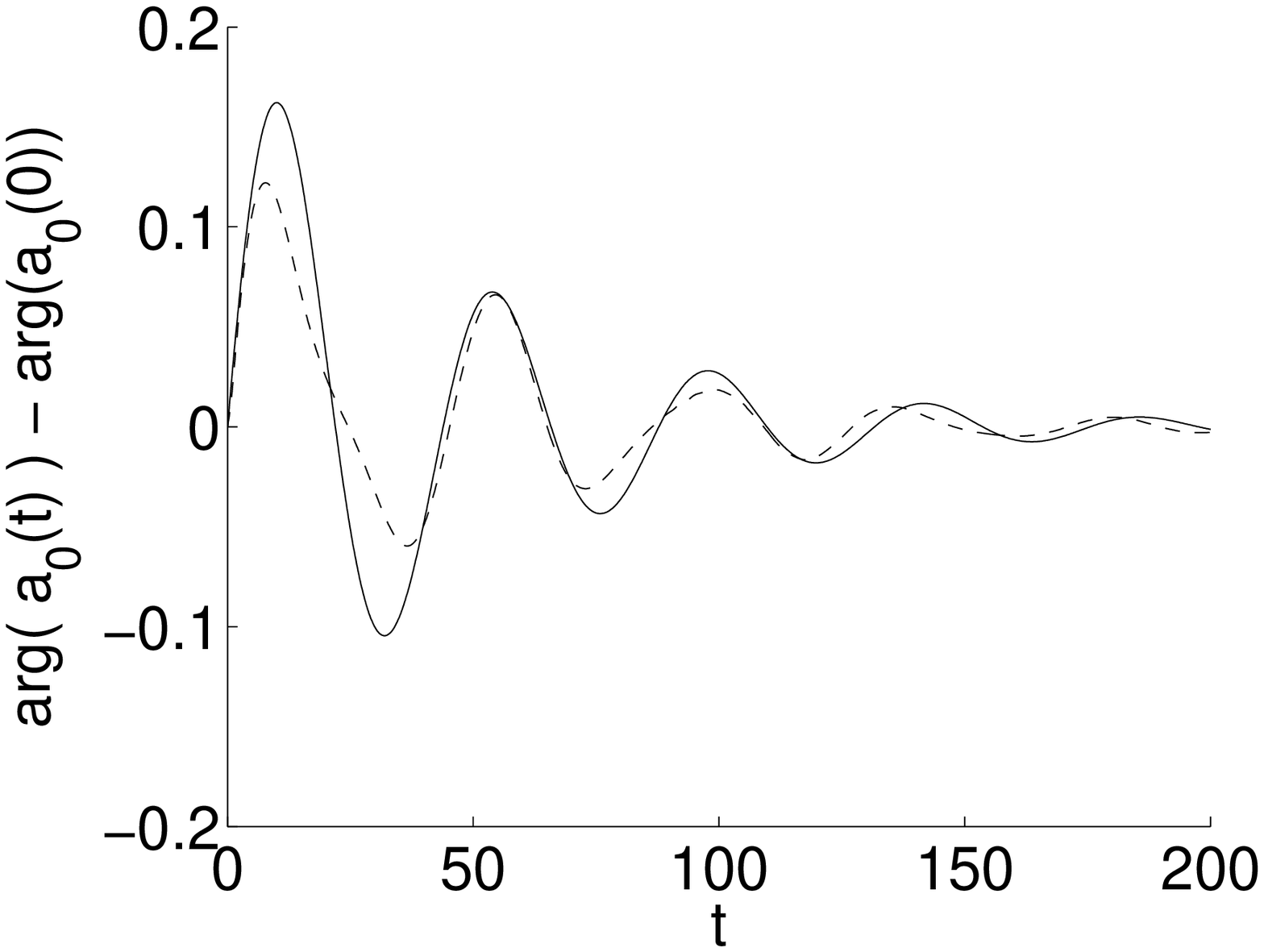}
\includegraphics[width=5.0cm]{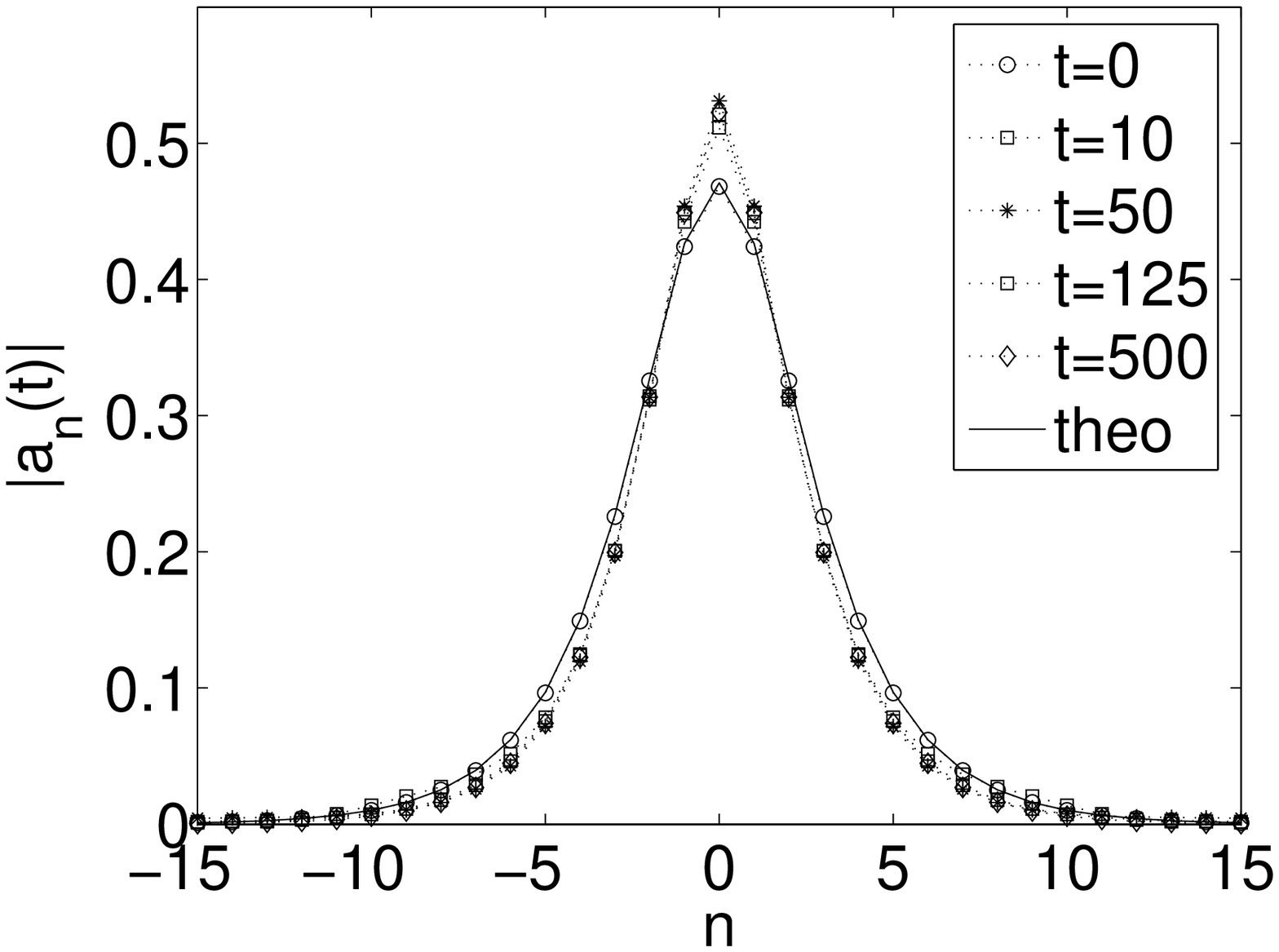}
\end{tabular}
\vspace*{-0.25in}
\end{center}
\caption{Here $\chi=1$,  $\omega=0.3$, $\delta =0.022$, $\gamma=0.02$.
The $T$ and $t$ scales coincide.
The initial condition is a soliton with $\sigma=\sigma_+$, $x_+=0$,
$\alpha=0$, and $\beta=\beta_+ - 0.02=0.43$.
We plot in the left picture the argument of $a_0(t)$ (dashed line).
The observed oscillations and damping are correctly predicted by the model.
The period is $2 \pi / \Omega_+=43.9$ and the exponential decay rate
is $\Gamma=0.02$ (solid line).
Note also, in the right picture, that the stationary profile is not exactly the sech predicted
by the AL theory, but a slightly deformed version.
\label{fig5} }
\end{figure}

When $\chi$ is large (say equal to $1$),
it is important to choose a suitable value $\omega$
so that the soliton parameter $\beta_+$ defined by
(\ref{eq:defbetapm2})
is small (more exactly, smaller than $0.5$).
Indeed, the theoretical
analysis based on the perturbed IST  is valid only in this case.
Furthermore, numerical simulations show that (\ref{eq:defbetapm2}) is not
a stationary point for larger values of $\beta_+$, which shows the fundamental
limitation in the perturbed IST.
Within this region of parameters,
the  soliton parameters $\beta$ and $\sigma$
oscillate with the frequency $
\Omega_+ $ given by (\ref{eq:dnls:omega+})
and they also experience an exponential decay with the rate $\Gamma$.
The oscillation period and damping rate are confirmed
by numerical simulations (Fig.~\ref{fig5}).

\begin{figure}
\begin{center}
\begin{tabular}{c}
\includegraphics[width=5.7cm]{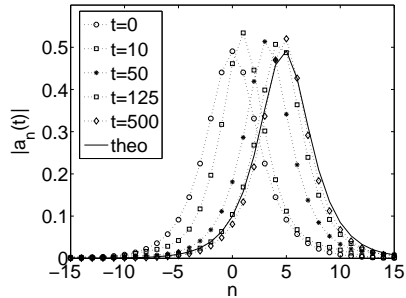}\end{tabular}
\vspace*{-0.25in}
\end{center}
\caption{Here $\chi=1$, $\omega=0.3$, $\delta =0.022$, $\gamma=0.02$.
The initial condition is a soliton with $\sigma=\sigma_+$, $x_+=0$,
$\alpha=0.1$, and
$\beta=\beta_+ =0.45$.
We plot the soliton profiles $|a_n(t)|$ at different times,
which exhibits the trapping of the moving soliton.
The solid line is the theoretical stable stationary soliton centered at $x_F = 4.82$
given by (\ref{eq:dnls:defxf}).
\label{fig6} }
\end{figure}

Moreover, as in the AL case,
the propagation of moving solitons is not supported,
as the soliton velocity decays exponentially to $0$.
If we denote by $\alpha_0$  the initial value of the parameter $\alpha$,
then the input soliton converges to its stationary form
centered at
\begin{equation}
\label{eq:dnls:defxf} x_F  = (1-\chi) \frac{ \sinh\beta_+ \sin
\alpha_0 }{\beta_+\Gamma} +\chi \frac{ \sinh^2 \beta_+ \sin
\alpha_0 }{\beta_+^2\cosh \beta_+\Gamma}.
\end{equation}
Notice that we have neglected higher-order term (in $\beta$) in
this expression, which is consistent with the previous hypotheses
and which gives a very accurate prediction for the final soliton
position (see Fig.~\ref{fig6}).

\section{Conclusion}
In this paper we have have investigated the existence and
stability properties of new types of bright discrete solitons in
discrete nonlinear Schr\"odinger-type models with damping and
strong rapid drive. Stable stationary solitons are exhibited in
the case of a general cubic nonlinearity. If the nonlinearity has
the special AL form, then stationary solitons,  moving solitons,
and periodic trains of solitons are found to be stable solutions
of the system. These results have been obtained by applying a
perturbed inverse scattering transform to the averaged equation
and confirmed by numerical simulations. This means that the
inverse scattering theory  is useful for probing the parameter
space and exhibiting interesting phenomena. One of the problems
that should be addressed for future consideration is the
mechanisms  responsible for the instablities of the periodic cn
and dn solutions for large drive, which seem to be different since
a chaotic instability appears first in the dn case, while new
patterns with different periods appear in the cn case.

\end{document}